# Transient Nature of Fast Relaxation in Metallic Glass

Leo Zella[1], Jaeyun Moon[2*], David Keffer[1], Takeshi Egami[1,2,3†]

[1]Department of Materials Science and Engineering, The University of Tennessee, Knoxville, Tennessee 37996, USA

[2]Materials Science and Technology Division, Oak Ridge National Laboratory, Oak Ridge, Tennessee 37831, USA

[3]Department of Physics and Astronomy, The University of Tennessee, Knoxville, Tennessee 37996, USA

* Electronic mail: moonj@ornl.gov, †Electronic mail: egami@utk.edu

## Abstract

Metallic glasses exhibit fast mechanical relaxations at temperatures well below the glass transition, one of which shows little variation with temperature known as nearly constant loss (NCL). Despite the important implications of this phenomenon to in aging and deformation, the origin of the relaxation is unclear. Through molecular dynamics simulations of a model metallic glass, $Cu_{64.5}Zr_{35.5}$, we implement dynamic mechanical analysis with system stress decomposed into atomic-level stresses to identify the group of atoms responsible for NCL. This work demonstrates that NCL relaxation is due to fully transient groups of atoms that become normal over picosecond timescales. They are spatially distributed throughout the glass and have no outstanding features, rather than defect-like as previously reported.

# 1. Introduction

Identifying atomic origins of various dynamic processes is crucial to understanding mechanisms of various properties in amorphous solids, such as aging, relaxation, and deformation [1–5]. Metallic glasses are especially useful to study because of a relatively simple disordered atomic structure due to their close packing not dominated by local chemical bonds [6,7]. The picture of relaxation process in metallic glass has continued to evolve over time and now includes many processes depending on their length and time scales including $\alpha$ relaxation [8], $\beta$ relaxation [9–11], fast $\beta$ relaxation [12,13], Boson peak dynamics [14], a low temperature relaxation peak (LTP) [15], and the subject of this paper, nearly constant loss (NCL)[16–18].

Dielectric loss measurements of glycerol and propylene carbonate by Lunkenheimer et al. [19] revealed "additional fast processes" between the boson peak and excess wing of the $\beta$ relaxation, which is now characterized as nearly constant loss (NCL). NCL has been subsequently observed in numerous other disordered systems such as polymers, ionic conductors, and metallic glass [18,20,21]. The transition from NCL to $\beta$ relaxation has been measured by dynamic mechanical analysis (DMA) and quasi-static tensile measurements in $La_{60}Ni_{15}Al_{25}$ bulk metallic glass (BMG), demonstrating the role of NCL in relationship to other processes [21]. The same experiment measured the activation energy and found it to be roughly half that of the $\beta$ relaxation which may indicate a more localized nature of relaxation [21]. The hypothesized origin of NCL is thought to be rattling or fast caged dynamics which may be described by mode coupling theory (MCT) [18,20,22,23].

Since dielectric loss measurements and dynamic mechanical analysis are macroscopic measures, and it is not possible to identify the microscopic dynamics from these measurements alone. Classical Molecular Dynamics (MD) investigations of $Cu_{65}Zr_{35}$ metallic glass investigated "fast processes", i.e. dynamics between boson peak and $\beta$ relaxation which includes NCL, and identified a 40 GHz rattling mode that resulted in 'string-like' rearrangements of 2-3 neighboring atoms to the rattling center [24]. However, the amount of contribution of rattling type motion to NCL is undetermined as there has been no systematic measure of all dynamics related to the NCL relaxation. Additionally, the time dependence of atoms involved during the relaxation has not been resolved. Furthermore, the definition and precise characterization of rattling atoms are missing. Thus, despite these prior attempts to gain understanding of the NCL relaxation, the origin and mechanism of NCL is unclear.

In this work, we systematically identify the transient groups of atoms responsible for NCL using the standard MD-DMA technique combined with atomic level stress calculations. Instead of focusing on specific dynamics such as rattling, we directly connect atoms causing loss modulus to the relaxation mechanism. We show that the NCL relaxation sites are transient and spatially homogenous. We examine the correlation of NCL relaxation with atomic level parameters often used in identifying defective sites, such as non-affine displacement, atomic-level stresses, atomic volume, and coordination number. We find no clear distinct features of atoms responsible for NCL, demonstrating the subtlety of the NCL mechanism.

## 2. Methods

### 2.1 Model Metallic Glass

Classical Molecular Dynamics simulations were performed using the Large-scale Massively Parallel Simulator (LAMMPS) [25] with a timestep of 1 fs. Periodic boundary conditions were imposed. As a prototypical metallic glass, a $Cu_{64.5}Zr_{35.5}$ glass with 4000 atoms and cubic side lengths of 39.9 Å was made by a typical melt-quench method. The Embedded atom method (EAM) potential was used to describe interatomic interactions [26]. First, the system was melted at 3000 K for 1 ns. It was then cooled down to 300 K with a cooling rate of 1 K ps$^{-1}$ in the NVT ensemble. Equilibration was done at 300 K under NPT ensemble with zero pressure for 100 ps.

### 2.2 Molecular Dynamics Dynamic Mechanical Analysis (MD-DMA)

By applying a sinusoidal strain to the system mimicking experimental DMA, the relaxation dynamics of the system is probed as in Ref. [27]. A sinusoidal shear strain was applied following $\epsilon(t) = \epsilon_A exp(i\omega)$, where $\omega = 2\pi/T_\omega$, with a maximum strain of $\epsilon_A = 1\%$ and a period of $T_\omega = 50$ ps. The system was in the linear elasticity regime throughout. In this paper, 30 cycles were done for a total production time of 1500 ps. We anticipate that our results are independent of system size and strain frequency as demonstrated by Yu et al. [28]. Storage and loss modulus are related to the sinusoidal amplitude and strain by $G' = Re(\sigma_A/\epsilon_A)$ and $G'' = Im(\sigma_A/\epsilon_A)$, respectively, where $\sigma_A$ is complex sinusoidal stress amplitude, $\epsilon_A$ is magnitude of strain, and $\delta$ is the phase shift [29]. In this work, we will focus on the phase shift $\delta$ as it indicates the magnitude of loss modulus. To study the transient nature of the NCL relaxation, we divided the 1.5 ns simulation into 15 blocks of 100ps each in length and calculated the phase shift of the entire system of 4,000 atoms for each block.

## 2.3 Atomic-Level Stress, frequency representation and partial stress

To gain an atomic perspective on the origin of the NCL, we calculated atomic-level stresses, which represent the first order response of an atom to a strain [30,31]. By partitioning the system stress into individual atomic response, we can describe the local stress environment of an atom. Pairwise atomic-level stress is given by

$$\sigma_i^{ab} = \frac{1}{2v_i} \sum_j \frac{1}{r_{ij}} \frac{d\phi}{dr_{ij}} r_{ij}^a r_{ij}^b + \frac{1}{2v_i} m_i V_i^a V_i^b \qquad (1)$$

Where $a$ and $b$ are Cartesian components, $v_i$ is the atomic-level volume, $\phi$ is the two-body potential, $r_{ij}^a$ is the $a$ component of the distance vector, $\boldsymbol{r_{ij}}$, between atoms $i$ and $j$, and $V_i^a$ is the $a$ component of the velocity of atom $i$, $\boldsymbol{V_i}$. This calculation is done in LAMMPS where both the kinetic energy term and pairwise term were used and accounts for periodic boundary conditions [32]. At such low temperature kinetic contribution is expected to be negligible. Atomic-level volume is calculated by Voronoi tessellation [33].

By transforming the time dependent atomic-level stress response into frequency space by Fourier transform, we identify the individual stress response amplitude and phase for each atom. The individual complex atom shear modulus is defined by $G = \sigma_i^{ab}(t)/\varepsilon(t)$, and phase shift by $\delta_i = \arctan(Im(G)/Re(G))$. In typical MD-DMA calculations the phase shift is limited to $[0, \frac{\pi}{2}]$. However, locally the response can be in an opposite direction to the strain. For this reason, we computed the atomic level phase shift from $[-\pi, \pi]$. The system stress response is thus partitioned amongst individual atoms and the system stress response sum is recovered by $\sigma_{xy}(t) = \sum_i \sigma_{A,i} \sin(\omega t + \delta_i)$ with $\sigma_{A,i}$ being the amplitude of atom, $i$, sinusoidal stress response. The system stress response sum is shown to be accurate in phase and amplitude (See Fig. S1 in Supplementary Materials).

To investigate a group of atoms that may be responsible for the loss modulus during sinusoidal straining, partial stress excluding the contributions from selected atoms is calculated. New phase shifts, $\delta_{partial}$, and amplitudes of the partial stress are then found. If the partial stress has a smaller phase shift after excluding the atoms of interest, this means that the excluded set of atoms was contributing strongly to the overall phase shift and thus the loss modulus of the system. In this work, a group of 5% of the atoms (200 atoms) were excluded to form the partial stress. Error bars are calculated by the bootstrap method [34].

## 3. Results

### 3.1 Atomic phase shift distribution and transient phase shift

The system-level stress as a function of time is shown in Fig. 1a in blue for a time range corresponding to the middle of the imposed strain. The overall stress response due to the oscillating strain is well fit by a sine function, and is similar to the strain, shown by a dashed curve in yellow. The phase shift in this case is small. In Fig. 1b, the atomic-level stress for a single representative atom is shown. The fluctuation around the best-fit sine function is much greater than that for the system, which was averaged over 4,000 atoms. However, the clear oscillatory nature is evident in the figure. Curves such as the one shown were collected for each atom. The phase shift and amplitude for either the system or individual atoms were extracted from the corresponding stress response via Fourier transform. An example Fourier transform for an individual atom is shown in Fig. 1c. The largest peak at 0.02 THz corresponds to the principal response to the strain, whereas other components are thermal and mechanical noise.

The calculation of individual atomic-level phase shift and amplitude allows for insight into the atomic-level viscoelastic response. Fig. 2a. shows a representative distribution of atomic level phase shift for all atoms from 700ps to 800ps which is in the middle of the sinusoidal straining and after equilibration of initial sinusoidal strain. The atomic level phase shifts have a wide distribution from $-\pi$ to $\pi$ and reveal complex viscoelastic response that exists in glass due to the lack of periodicity. Based on analogy with the bulk experiment, a phase shift of $\delta_{atom} = 0$ corresponds to an elastic response. The distribution of atomic-level phase shifts shows a peak in the elastic response with local maximum at $\delta_{atom} = -\pi, \pi$ and minima at $\delta_{atom} = -\frac{\pi}{2}, \frac{\pi}{2}$.

Taking the partial stress and excluding 5% of the atoms from certain regions shows the role of each group of atoms on the phase shift and thus energy dissipation. These partial stress calculations centered at $\delta_{atom} = -\frac{\pi}{2}, 0, \frac{\pi}{2}, \pi, -\pi$ from the atomic phase distribution, shown in Fig. 2a with different colors. Seen in Fig. 2b are the calculated phases shifts from partial stress over the 15-time blocks of 100 ps arranged sequentially. It is clearly shown that the removal of 200 $\delta_{atom} \approx 0$ atoms, represented by orange crosses have little to no impact on the system phase shift as expected. These atoms are mostly elastic and contribute to the central peak in the distribution. On the opposite end, removal of 200 $\delta_{atom} \approx \frac{\pi}{2}$ atoms, depicted by green squares has a significant impact on the phase shift of the partial stress. Interestingly, the exclusion of the group of atoms with $\delta_{atom} \approx -\frac{\pi}{2}$ cause the partial stress phase shift to increase. The 'out of phase' atoms with $\delta_{atom} \approx \pi, -\pi$ have minimal impact on system phase shift. The group composed of 5% of atoms with phase shift around $\delta_{atom} \approx \frac{\pi}{2}$ as those that have the greatest impact on reducing the partial stress phase shift and thus can be identified as the main contributors to NCL. To confirm our results of phase shift reduction, the loss modulus was also calculated using the same groups of excluded atoms and is seen in Fig. 2c. We see the same trends as in Fig. 2b. These results show that the atomic-level responses are strongly heterogeneous at any instant in time. Whereas a large portion of atoms show basically elastic response, a fraction of atoms behave in a varied complex manner. The observed system-level energy loss results from contributions of individual atoms undergoing both energy loss and energy gain.

### 3.2 Transient nature of atomic characteristics

We now examine the temporal behavior of atoms responsible for NCL relaxation. Fig. 3 shows how the nature of a group of atoms which had the initial phase shift, $\delta_{atom} \approx \frac{\pi}{2}$, evolves with time. As seen here atoms causing NCL during 700-800 ps have a distribution shown in Fig. 3a, but in the next 100 ps cycle (Fig. 3b) they show widely distributed phase shifts, which resemble the system distribution except for a high phase group between $\delta_{atom} \approx [\frac{\pi}{2}, \pi]$. This rapid change of the phase distribution shows that after participating in NCL action atoms move to different states having a normal local dynamic response to the system strain. Finally, these atoms are tracked at the end of the simulation in Fig. 3c where the distribution still resembles the normal system distribution. This surprising result shows that characterizing a single group of atoms at a fixed time as the origin of NCL relaxation is misleading and the transient nature is an important factor.

A spatial visualization of the NCL causing atoms is seen in Fig. 4. During 700-800 ps and from the next two cycles 800-900 ps shows that there is very little overlap in the groups of atoms causing the NCL relaxation. The spatial distribution appears uniform with no local aggregation as opposed to string-like distributions seen related to $\beta$ relaxation [11,35]. There also does not appear to be a spatial correlation between regions of previous relaxation and subsequent atoms responsible for NCL. The spatial visualization confirms our results that the NCL relaxation evolves in time by involving different groups of atoms showing a stochastic behavior.

**3.3 Subtle nature of Nearly Constant Loss (NCL)**

We performed further analysis to characterize the intrinsic nature of atoms responsible for NCL relaxation in terms of various factors which have been used in literature, including non-affine displacement, atomic-level shear stress, atomic-level volume, and coordination number, which are summarized in Table 1. In a defect view, the NCL relaxation may be caused by local regions of low density, high stresses or over/under coordination of neighbors. By calculating the averages and variance of atomic properties for the atoms causing NCL at the beginning and end of the time blocks and comparing to the system averages we try to ascertain if local property changes are related to NCL relaxation. We find that the atomic properties of the set of atoms causing NCL versus the system to show negligible differences.

Prior works investigating structural relaxation during sinusoidal straining have calculated non-affine mean squared displacement to determine mobility [15,28] and relate the mechanism of structural relaxation to the fastest non-affine atoms. We, therefore, calculated the non-affine displacements for every 100 ps which corresponds to 2 cycles of sinusoidal straining to understand the mobility of NCL causing atoms with $\delta_{atom} \approx \frac{\pi}{2}$. The results of the distribution are shown in Fig. 5 for all atoms in the system compared to atoms causing NCL relaxation. In the logarithmic scale on the y-axis, we are able to observe that NCL atoms have a tail of slightly larger non-affine displacements and overall have a broad distribution of non-affine displacement but on average are not more mobile as seen in the non-affine displacement value in Table 1.

4. Discussion

Understanding the NCL relaxation is difficult due to lack of characteristic peak in temperature dependence. Because there are no direct experimental measurements of NCL relaxation length scale to shed light on the nature of the dynamics, simulations [24] and theory [18] are the only tools to provide insight into the atomic level dynamics of NCL. To identify atoms responsible for relaxation during MD-DMA, we calculated the DMA phase shift for each atom and identified the group of atoms with an atomic-level phase shift of $\delta_{atom} \approx \frac{\pi}{2}$ to be primarily responsible for NCL. On the other hand, the group of atoms at $\delta_{atom} \approx -\frac{\pi}{2}$ have a 'stabilizing' type effect in that their exclusion from partial stress resulted in an increase in the apparent mechanical loss. These atoms must have had stored energy prior to deformation and released the energy during deformation, thus resulting in negative phase shift. Atoms with no mechanical loss which are in the central peak of atomic phase shift in Fig. 2a account for roughly 80% of all atoms at any given time. The fraction of the atoms which have a high atomic-level phase shift representing inelastic response, 20%, is consistent with measurements of liquidlike atoms measured by x-ray scattering [36].

Having identified the atoms causing NCL as those with an atomic level phase shift of $\delta_{atom} \approx \frac{\pi}{2}$, we examined the nature of the atoms which show high levels of atomic-level viscoelastic loss. Whereas shear transformation zones (STZ) have been attributed to loosely packed regions [37] and loosely packed regions are also thought to be related to fast relaxation processes in $Cu_{65}Zr_{35}$ metallic glass [24], there are no differences in the distribution of atomic-level properties of the system vs. viscous atoms causing NCL. The lack of difference of atomic-level volumes suggests that NCL is not due to local free volume fluctuations that have previously been related to structural relaxation [38,39] or due to local structural defects that can be measured by atomic-level stress [30]. The mean squared non-affine displacements measured as seen in Fig. 5 show the comparison between all atoms and atoms causing NCL which are only slightly more mobile. The low temperature peak (LTP) in the temperature dependence of mechanical loss is linked to "reversible atoms (RAs)" with maximum $u'_{max} \in [1\ \text{Å}, 2\ \text{Å})$ [40] and Yu et al. [41] related "faster atoms" with displacements larger then $u\ast = {r_0}/{2} = 1.4\ \text{Å}$; however, in our case no such groupings of mobility were found to be linked to the NCL relaxation from the point of view of phase shift. There was no evidence that fast-moving atoms were solely responsible for NCL. These observations suggest a different mechanism for NCL relaxation compared to the proposed mechanism of $\beta$-relaxation where string-like motion by a small proportion of highly mobile atoms causes structural relaxation [10].

An important characteristic of NCL is its transient nature. We found that after mechanical loss event the atoms moved to states having a 'normal' response as shown by the rapid change in Fig 3. Our finding is consistent with the observation that after deformation event atoms lose their memory of the initial state [42]. Along the pathway of moving from a saddle point to a valley in the potential energy landscape for liquids and glasses, saddle states show universal melting behavior wiping out prior thermal history of atomic configurations [42]. Spatial visualization done in Fig. 5 shows the groups of atoms causing NCL in two adjacent time segments. The lack of spatial overlap demonstrates no correlation with previous regions of NCL relaxation and is consistent with the idea of loss of thermal history. This observation is also consistent with the transient nature of STZ emphasized by Langer [44]. STZ is a transient state which disappears after the deformation event, and memory is retained only in the long-range back-stress field around it.

The technique of set-to-point has been used in previous works where certain atoms are made to deform affinely [15,43]. However, this technique interferes with local dynamics. By choosing certain groups of atoms to behave differently, this approach cannot be utilized for probing transient relaxation where atoms are only involved in the relaxation during some cycles of sinusoidal straining. Our technique of measuring atomic level loss through the Fourier transform of the atomic level stress thus represents a non-invasive method of directly identifying atoms responsible for mechanical loss. The NCL relaxation appears to be qualitatively and quantitatively different from known $\beta$-relaxation mechanisms. Therefore, if the observations made here to NCL apply also to $\beta$-relaxation or not is yet to be examined.

5. Conclusions

By utilizing the unique combination of atomic level stress and MD-DMA, we directly identified the group of atoms responsible for the NCL relaxation and have revealed the surprising transient nature of the microscopic mechanism. The diverse atomic level viscoelastic response identifies four groups of atoms that participate in the system viscoelastic response with one group primarily contributing to NCL relaxation. Our results demonstrate that NCL is due to a population of atoms with $\delta_{atom} \approx \frac{\pi}{2}$ which are randomly distributed in space. After participating in NCL, the $\delta_{atom} \approx \frac{\pi}{2}$ atoms move to the states with a phase shift distribution resembling the system in average, suggesting that these atoms no longer belong to the $\delta_{atom} \approx \frac{\pi}{2}$ group. This observation is supported by evidence of saddle state melting behavior wiping the thermal history. The atoms belonging to the $\delta_{atom} \approx \frac{\pi}{2}$ group, however, show no fundamental difference in local atomic parameters, such as atomic level shear stress, coordination number, and atomic volume. This indicates that NCL does not occur at local structural defects, such as free volume fluctuations, but it is a totally stochastic event. Whether this observation applies also to other mechanical deformation events, such as $\beta$-relaxation, remains to be seen.

**Acknowledgment**

This research was supported by the U.S. Department of Energy, Office of Science, Basic Energy Sciences, Materials Science and Engineering Division. This research used resources of the National Energy Research Scientific Computing Center (NERSC), a U.S. Department of Energy Office of Science User Facility located at Lawrence Berkeley National Laboratory, operated under Contract No. DE-AC02-05CH11231 using NERSC award BES-ERCAP0017758. We thank H.B. Yu for discussion on the method for MD-DMA.

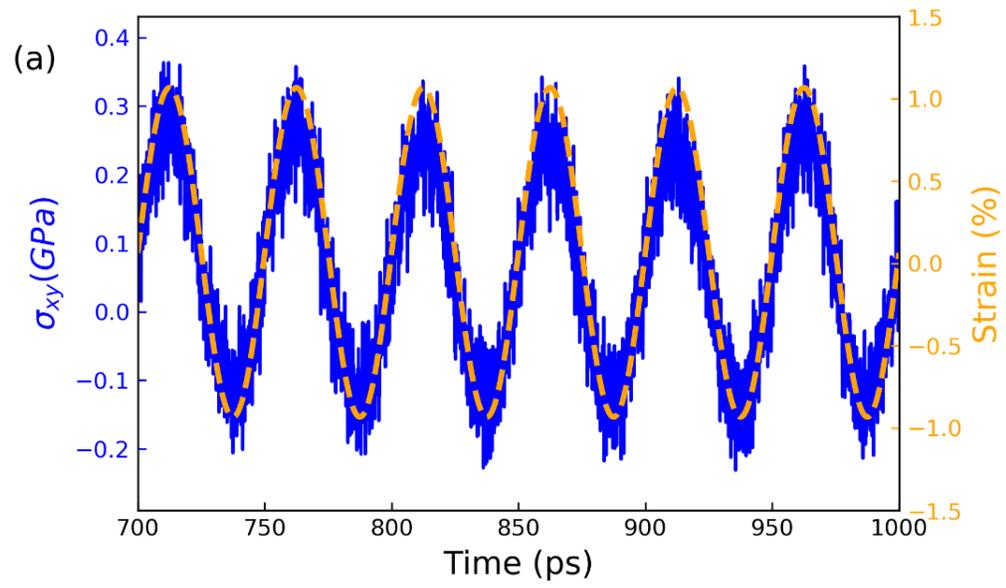

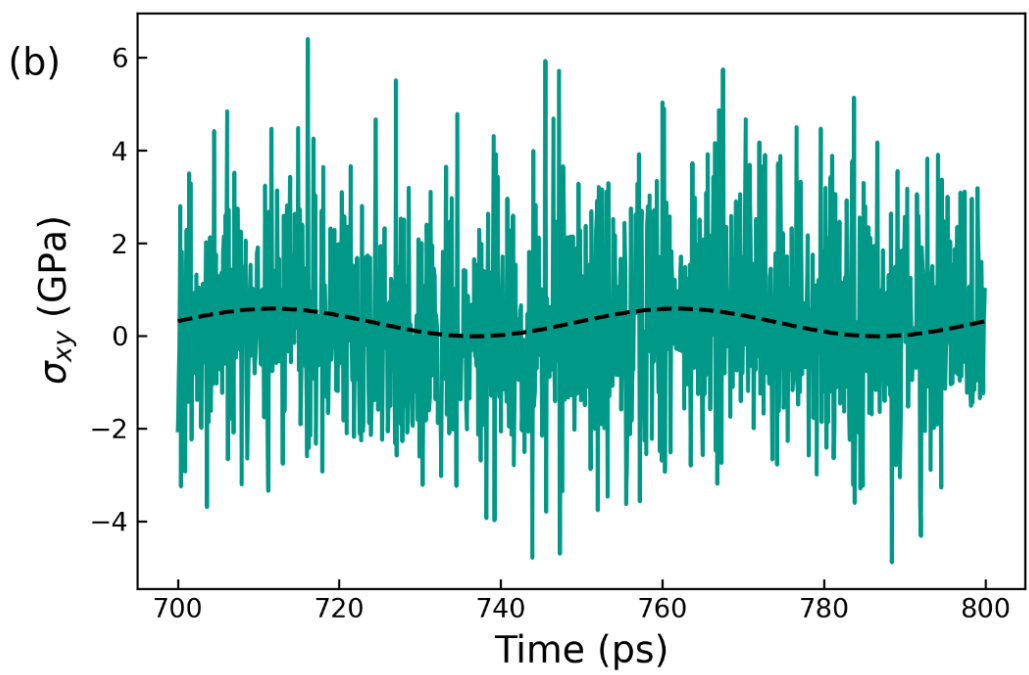

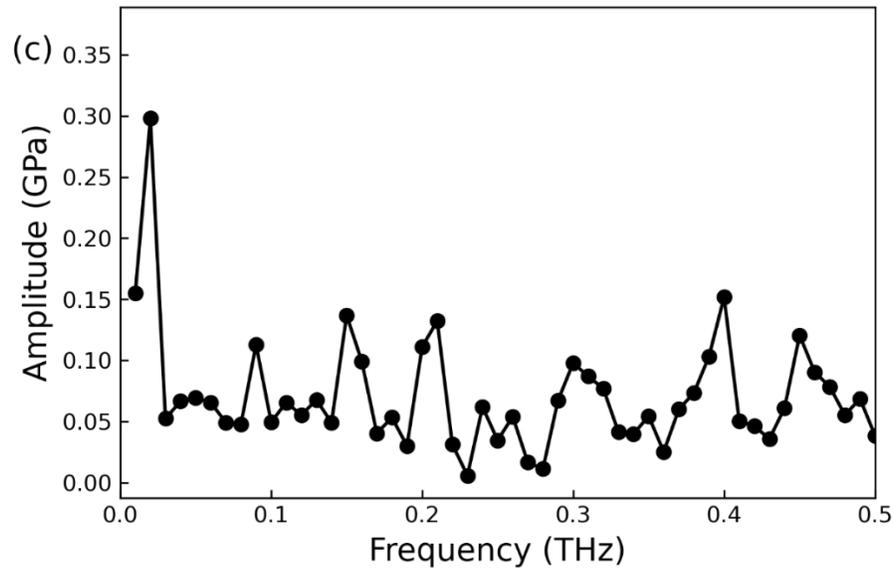

Fig. 1. (a) Shear stress and strain over 3 cycles of sinusoidal straining (b) Atomic level shear stress (teal) of an individual atom with sinusoidal fit (black dashed) calculated by Fourier transform. (c) Atomic level shear stress amplitude vs. frequency from Fourier transform.

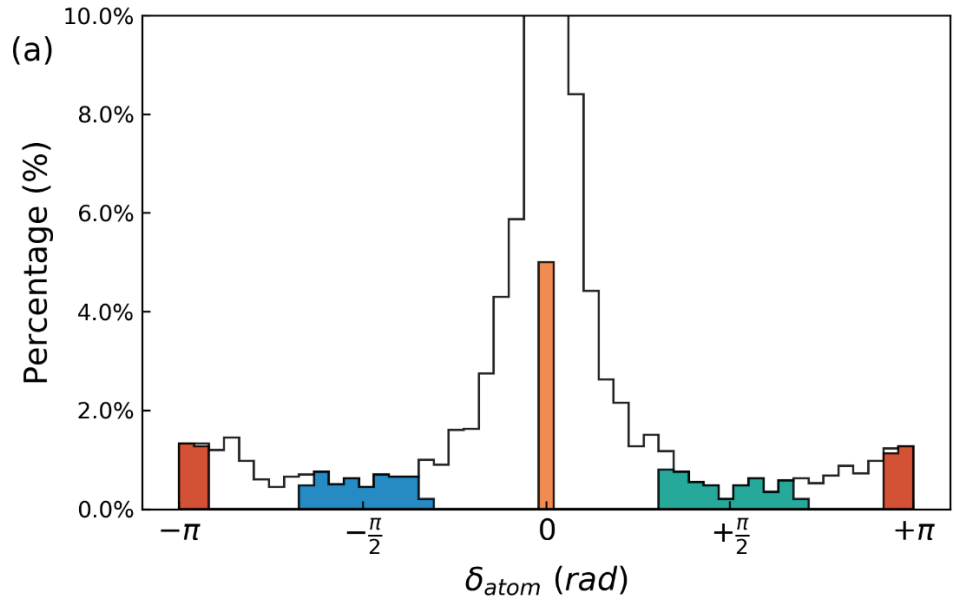

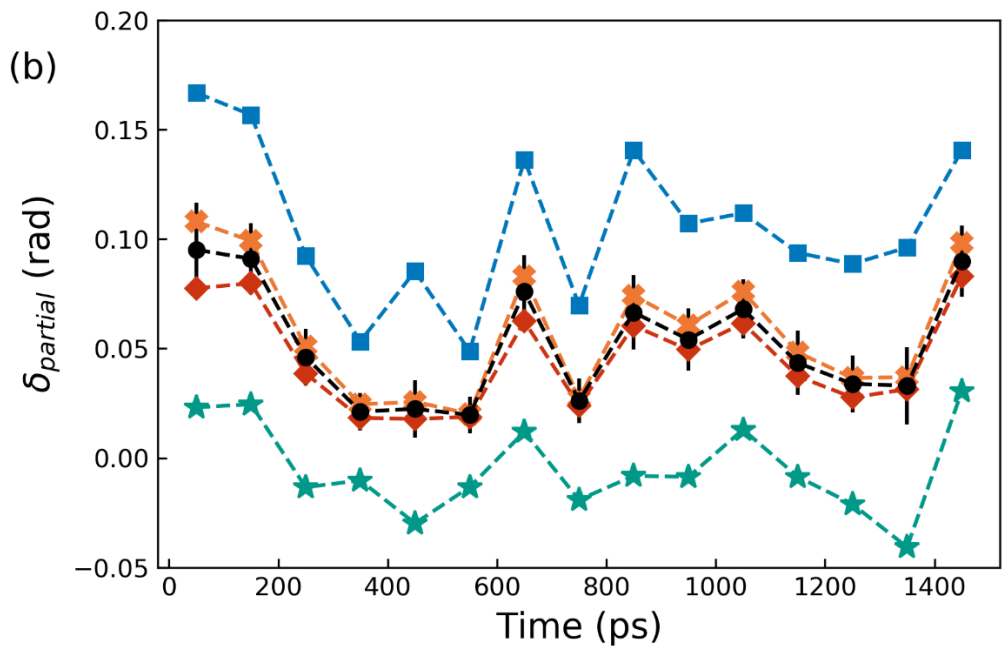

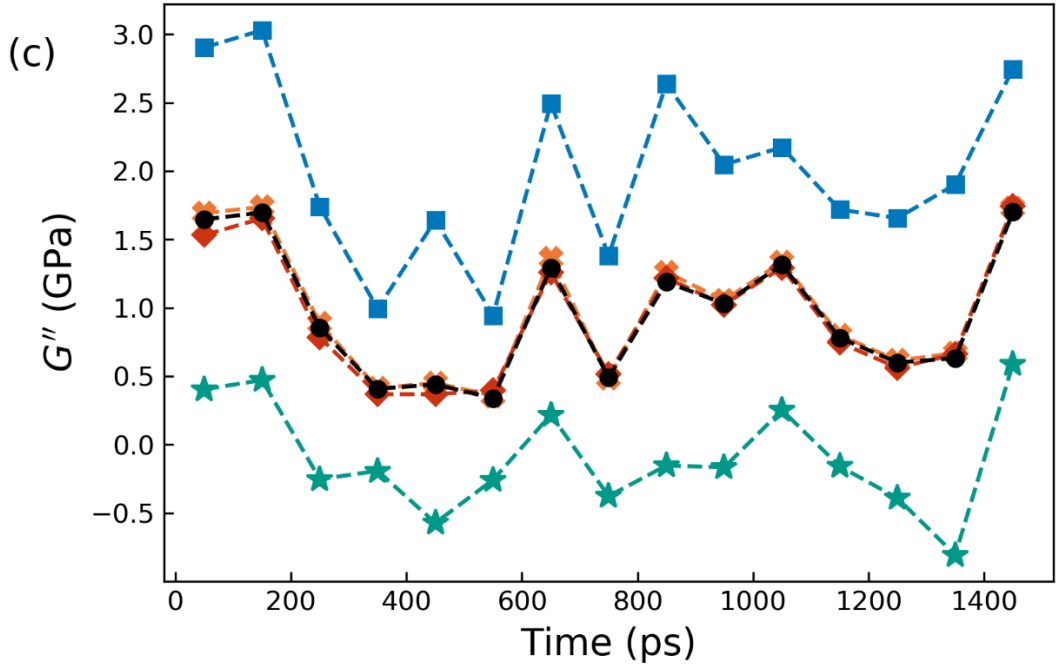

Fig. 2. (a) Representative phase shift distribution of all atoms during 700-800ps with shaded regions of interest that were examined. Nearly elastic atoms make up the central peak, the small populations of atoms at $\delta_{atom} \approx -\frac{\pi}{2}, \frac{\pi}{2}$ are shown to have a large impact on the viscoelastic response of the system (b) Partial phase shift from excluding differing groups of atoms by atom phase. Stars: $\delta_{atom} \approx \frac{\pi}{2}$, Circles: System phase shift with no reduction. Cross marker: $\delta_{atom} \approx 0$, Diamond marker: $\delta_{atom} \approx -\pi, \pi$, Squares: $\delta_{atom} \approx -\frac{\pi}{2}$. Atoms around $\delta_{atom} \approx \frac{\pi}{2}$ clearly have the greatest contribution to phase shift. (c) Loss modulus of respective groups after calculating partial phase shift. The same trend is observed as in the reduced phase shift.

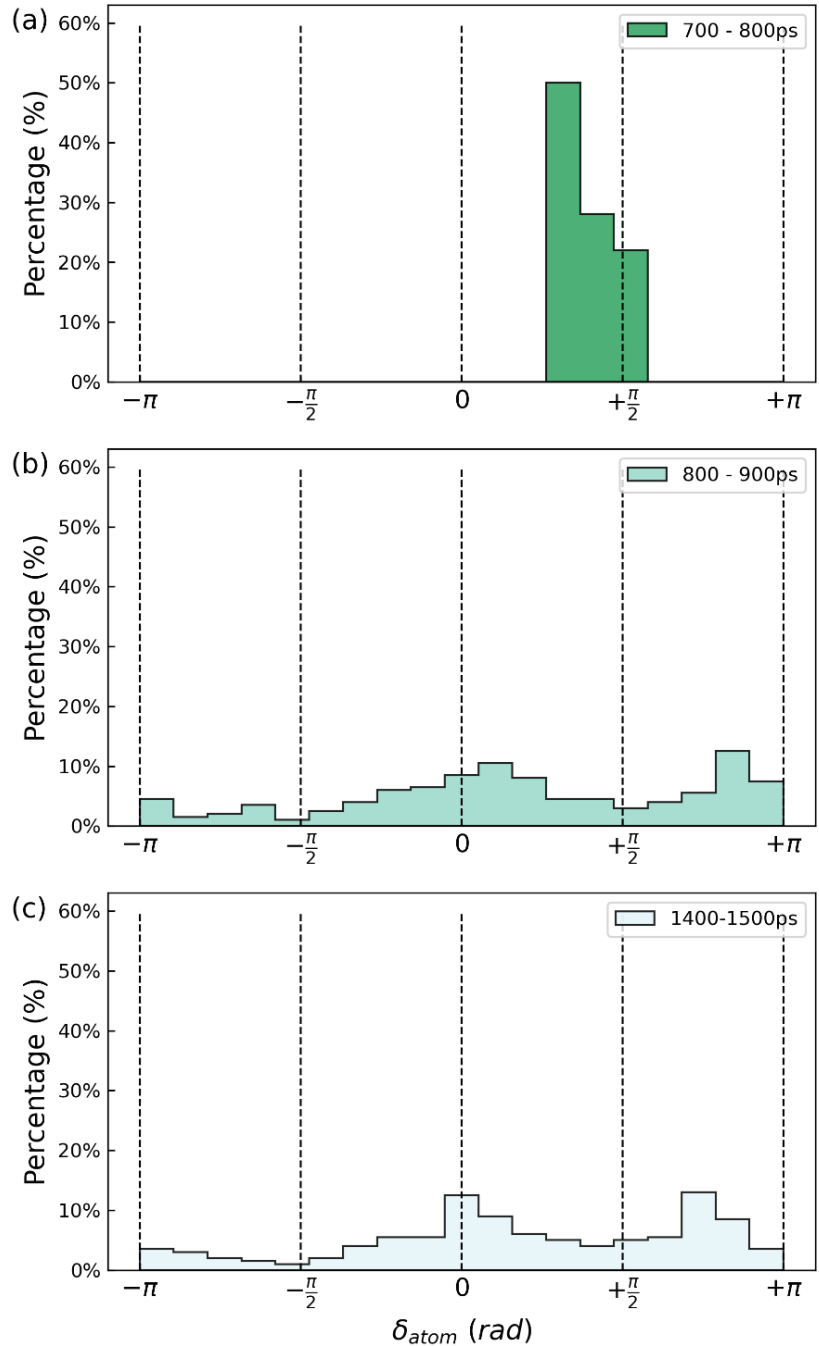

Fig. 3. The fast decay of the atomic phase distribution of NCL atoms from 700-800ps. (a) Phase shift distribution of NCL causing atoms during 700ps-800ps. (b) Phase shift distribution of same set of atoms during 800ps-900ps. (c) Phase shift distribution of same set of atoms in last 100ps of simulation. Once NCL atoms are involved in a relaxation the distribution quickly decays to resemble representative phase shift distribution with a high phase group of atoms.

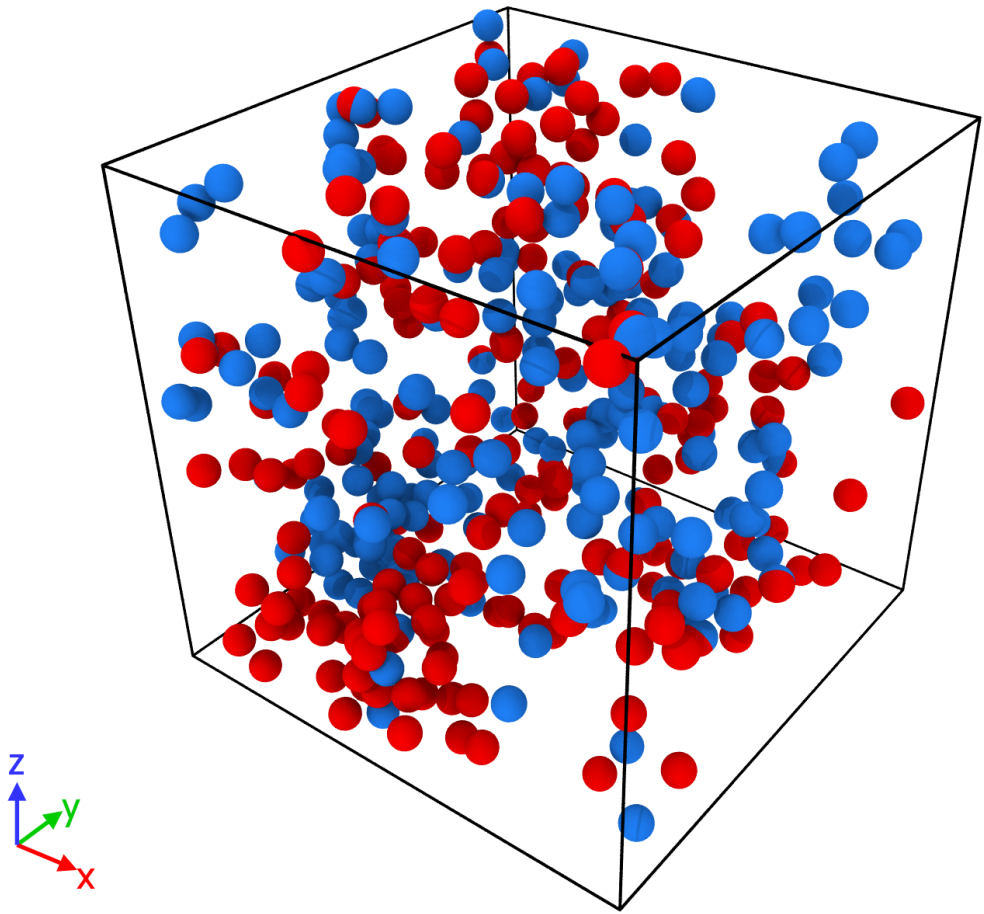

Fig. 4. Visualization of 200 NCL atoms from 2-time chunks. Blue: NCL causing atoms 700-800 ps. Red: NCL causing atom 800-900 ps. There is no obvious evidence of spatial aggregation but a homogenous distribution.

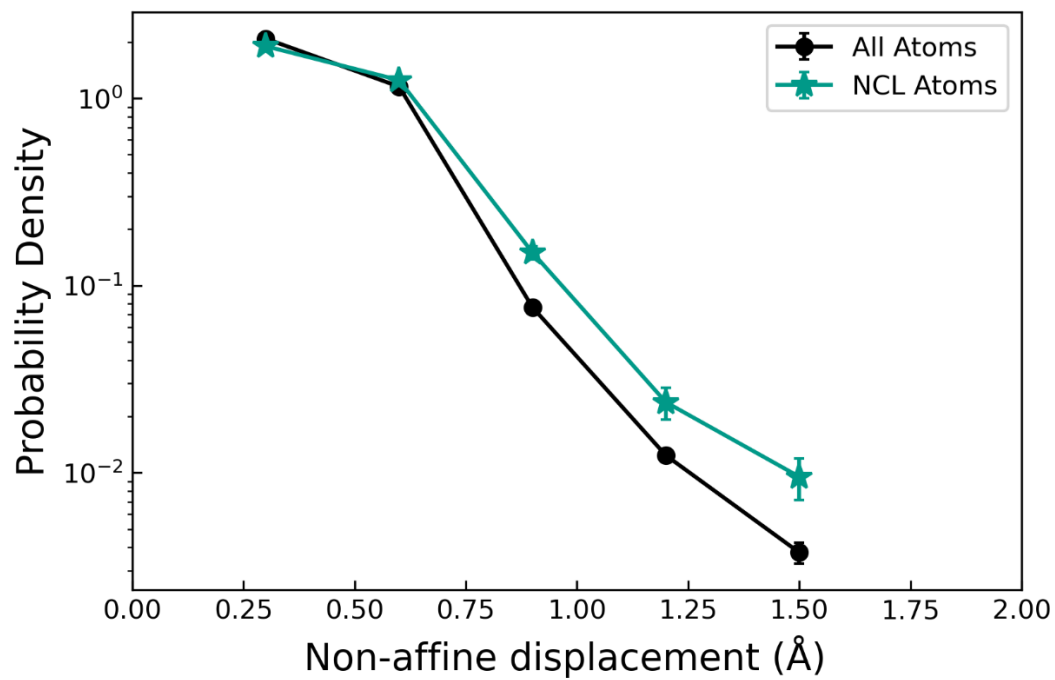

Fig. 5. Circles: Probability density of non-affine displacement for all atoms. Stars: Probability density of non-affine displacement for NCL atoms. NCL atoms have a wide range of displacements with slightly higher mobility on average. No evidence of preferred jumps or ranges of displacement. Error bars calculated by standard deviation of distribution.

Table 1. Variance and averages for $\sigma_{xy}$: atomic level stress shear, $N_c$: coordination number, $v_i$: atomic level volume. There are shown to be no substantive difference between NCL properties and system ensemble. Error calculated by block averaging.

|  | NCL Atom Group | System |
| --- | --- | --- |
| $Var[\sigma_{xy}](GPa^2)$ | 12.04 ± 0.55 | 11.54 ± 0.06 |
| $Var[N_c]$ | 1.59 ± 0.15 | 1.6 ± 0.0 |
| $<N_c>$ | 11.24 ± 0.09 | 11.35 ± 0.0 |
| $Var[V_{atom}](Å^6)$ | 3.28 ± 0.25 | 3.37 ± 0.01 |
| $<V_{atom}>(Å^3)$ | 15.88 ± 0.13 | 15.89 ± 0.0 |
| $<D(non-affine)>(Å)$ | 0.30 ± 0.17 | 0.28 ± 0.15 |

# SUPPLEMENTARY MATERIALS FOR

# "Transient Atomic Nature of Nearly Constant Loss in Metallic Glass"

Leo Zella[1], Jaeyun Moon[2*], David Keffer[1], Takeshi Egami[1,2,3†]

[1]Department of Materials Science and Engineering, The University of Tennessee, Knoxville, Tennessee 37996, USA

[2]Materials Science and Technology Division, Oak Ridge National Laboratory, Oak Ridge, Tennessee 37831, USA

[3]Department of Physics and Astronomy, The University of Tennessee, Knoxville, Tennessee 37996, USA

* Electronic mail: moonj@ornl.gov, †Electronic mail: egami@utk.edu

**Section I: Decomposition of system pressure into individual atom stresses during sinusoidal strain**

Using the Fourier transform individual phase and amplitudes for each atom may be calculated and the sum is compared to the system stress response. It is shown in Fig. S1 to be identical within error of the overall amplitude and phase shift validating the method.

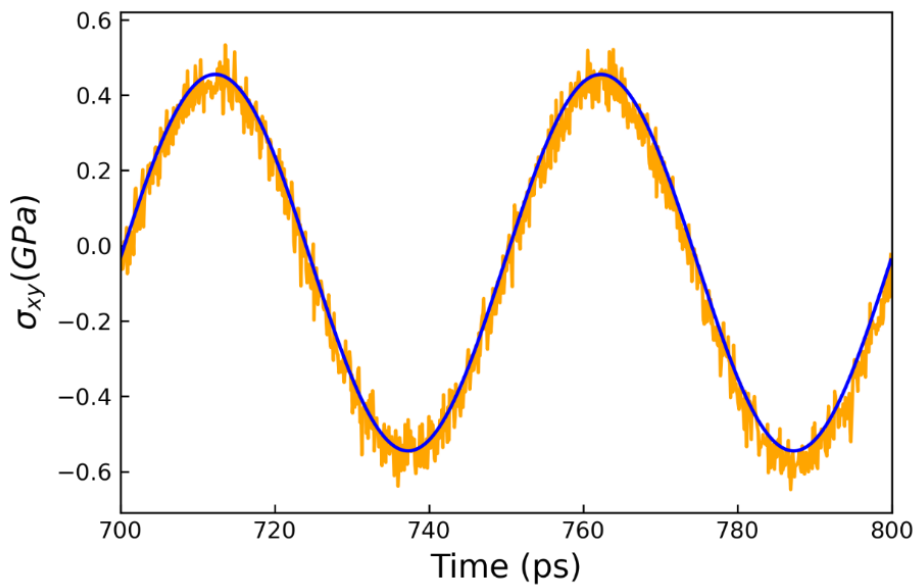

Fig. S1. Blue: System stress response. Orange: Sum of all atoms phase and amplitudes. System stress response from individual atom Fourier Transform method is shown to be accurate in phase and amplitude when compared to system stress response.